Pressure Effect on the Superconducting and Magnetic Transitions of the Superconducting Ferromagnet $RuSr_2GdCu_2O_8$

B. Lorenz [1], R. L. Meng [1], Y. Y. Xue [1], C. W. Chu [1,2,3]

[1] Texas Center for Superconductivity and Department of Physics, University of Houston, Houston, TX 77204-5932

[2] Lawrence Berkeley National Laboratory, 1 Cyclotron Road, Berkeley, CA 94720

[3] Hong Kong University of Science and Technology, Hong Kong, China

**Abstract**

The superconducting ferromagnet $RuSr_2GdCu_2O_8$ was investigated at high pressure. The intra-grain superconducting transition temperature, $T_c$, is resolved in ac-susceptibility as well as resistivity measurements. It is shown that the pressure shift of $T_c$ is much smaller than that of other high-$T_c$ compounds in a similar doping state. In contrast, the ferromagnetic transition temperature, $T_m$, increases with pressure at a relative rate that is about twice as large as that of $T_c$. The high-pressure data indicate a possible competition of the ferromagnetic and superconducting states in $RuSr_2GdCu_2O_8$.

1. **Introduction**

The coexistence of ferromagnetism and superconductivity in $RuSr_2LnCu_2O_8$ (Ru-1212) and $RuSr_2(Ln_{1+x}Ce_{1-x})Cu_2O_{10-\delta}$ (Ru-1222), where Ln = Gd, Eu, has attracted increasing attention after Ono [1] as well as Bauernfeind et al. [2] succeeded to synthesize this new class of ruthenium-copper oxides and demonstrated the existence of superconductivity in the Ce-doped Ru-1222. In the latter compound it was shown that weak ferromagnetism (ascribed to secondary phases by Ono [1]) apparently coexists with superconductivity [3]. The existence of superconductivity in the otherwise ferromagnetic $RuSr_2LnCu_2O_8$ was a matter of discussion [4] and could be confirmed only recently [5,6,7]. The physical nature of the superconducting and ferromagnetic states is far from being understood. Various reports on $RuSr_2LnCu_2O_8$ came to different conclusions. The spectrum of published data extends from non-superconducting samples [4] to samples showing zero resistance transition and zero field cooled (zfc) diamagnetic signal (but no field cooled (fc) or Meissner signal) [6,7,8] and eventually samples with a fc diamagnetic signal in the magnetic susceptibility [9] which is small and appears only at very small fields.

The investigation of the intrinsic superconducting properties of $RuSr_2LnCu_2O_8$ is further complicated by the fact that all samples investigated so far are polycrystalline ceramic pellets showing weak Josephson like inter-grain coupling in the superconducting state [7, 10]. The diamagnetic signal below the intra-grain superconducting transition temperature, $T_c$, is extremely small due to a large penetration depth [11] and may be masked by the huge diamagnetic shielding signal below the inter-grain phase-lock temperature, $T_p$, which is about 10 K lower. The assignment of the two transitions at $T_c$ and $T_p$ to intra- and inter-grain superconducting transitions, respectively, was recently proven in an investigation of the ac-susceptibility of sorted powders of $RuSr_2GdCu_2O_8$ [11,12]. The inter-grain diamagnetic signal systematically decreased with the particle size and completely disappeared for powders with an average particle size comparable to the grain size. The intra-grain diamagnetic signal remains constant until the particle size becomes smaller than the grain dimension and before the signal decreases with further reduction in particle size. The possible existence of microdomains inside the grains [11,12] has been suggested. The typical features of the multiple superconducting transition in $RuSr_2GdCu_2O_8$ are also reflected in the resistivity transition. The transport measurements reveal a rather broad superconducting transition that can be decomposed into two steps. Although this broadening of the resistive transition was already observed in early transport data [13] its interpretation as intra- and inter-grain superconducting transitions was discussed in more detail only recently [7,10,12]. It is essentially important to resolve the intra-grain superconducting properties since they reflect more about the intrinsic nature of the superconducting state and are independent of the granular structure of ceramic compounds.

The lattice structure of $RuSr_2LnCu_2O_8$ is similar to that of $YBa_2Cu_3O_{7-\delta}$ (YBCO) and is derived from the YBCO structure by replacing Y with Ln, Ba with Sr, and the CuO chains with a $RuO_2$ plane, respectively. However, unlike in the YBCO compound the oxygen content (and hence the doping level) cannot be changed at will in $RuSr_2LnCu_2O_8$. Thermogravimetric experiments [6] as well as measurements of the thermoelectric power [10] indicate that the oxygen content is close to 8 and cannot be changed by annealing in $O_2$ or other inert gases or by varying the conditions of synthesis. Because of its structural similarity the superconducting state of $RuSr_2LnCu_2O_8$ was frequently compared with that of under-doped YBCO (with similar transition temperature). In fact, the low carrier density of $n_h \approx 0.1$ holes/Cu deducted from transport measurements [14] and the room temperature Seebeck coefficient (60 to 70 µV/K) and its temperature dependence [10] are typical for an under-doped high-$T_c$ compound. Oxygen nuclear magnetic resonance studies [15] provided further evidence that Ru-1212 is similar to a very under-doped cuprate. The major difference between the YBCO and Ru-1212 systems is the magnetic order of the Ru-spins observed in the latter compound. Neutron scattering studies have shown that the Ru-spins order antiferromagnetically (G-type) below $T_m \approx 130$ K with a small ferromagnetic component of not more than 0.1 Bohr magneton even in the presence of a field of 1 T [16]. This ferromagnetic component is easily detected in dc- or ac-susceptibility measurements.

The coexistence of ferromagnetic order and superconductivity in the Ru-1212 and Ru-1222 compounds raises the question how these two antagonistic states of matter can accommodate each other. Do both states coexist with no mutual interference or is there a competition between superconducting and magnetic order? Based on muon spin rotation experiments it was suggested that the magnetic moments are not affected by the appearance of superconductivity in Ru-1212 below 45 K [5]. However, experiments on chemical substitution (doping) of Ru-1212 indicate that $T_c$ and $T_m$ are affected in an opposite way. Decreasing the hole density by partially replacing Gd with Ce [17] or Sr with La [18] results in a decrease of $T_c$ and an ultimate suppression of superconductivity and an increase of the ferromagnetic transition temperature. The magnetic transition of non-superconducting Ru-1212 was also shown to be a few degrees higher than $T_m$ of superconducting samples [19]. These results are indicative of a strong competition between superconductivity and magnetism. Chemical substitution usually affects several parameters at once. Besides the change of carrier density it may introduce disorder, reduce the magnetic coupling in the $RuO_2$ layers and cause changes of the microstructure of the sample.

We, therefore, decided to investigate the effect of hydrostatic pressure on the superconducting and magnetic phases. Pressure will not change the chemical composition but it is known from different high-$T_c$ compounds to increase $T_c$ with a typical coefficient of $dT_c/dp \approx 3…4$ K/GPa in the underdoped region [20,21,22]. The relative pressure coefficients of the intra-grain $T_c$ and of $T_m$ will be compared and discussed in the context of a competition of the superconducting and magnetic phases in $RuSr_2GdCu_2O_8$.

## 2. Sample preparation and experimental setup

Ceramic samples with a nominal composition $RuSr_2GdCu_2O_8$ were prepared by solid-state reaction techniques. The starting materials $RuO_2$, $Gd_2O_3$, $SrCO_3$, and $CuO$ were preheated at 600-800 ºC for 12 hours before used. The thoroughly mixed powder with cation ratio Ru:Sr:Gd:Cu = 1:2:1:2 was calcined at 960 ºC for 16 hours. The material was ground, compacted and subjected to additional sintering steps (10 to 24 hours each step) at successively increasing temperatures between 1015 and 1060 ºC. The sample preparation process was finished by long-term sintering (10 days) in oxygen atmosphere at 1065 ºC.

The magnetic and superconducting transitions of $RuSr_2GdCu_2O_8$ were investigated by ac-susceptibility and resistivity measurements at pressures up to 2 GPa. A dual coil system was mounted to the sample and four wires were attached for resistance measurements using indium pads. Resistivity and ac-susceptibility were measured simultaneously employing the resistance bridge (LR700, Linear Research). Pressure was generated in a beryllium-copper piston cylinder clamp. The sample was mounted in a Teflon container filled with a 1:1 mixture of Fluorinert FC70 and FC77 as a hydrostatic pressure transmitting medium. The pressure was measured insitu at 7 K by monitoring the shift of the superconducting $T_c$ of a high purity (99.9999 %) lead manometer. The temperature above 45 K was measured by a thermocouple inside the Teflon container and, at low temperatures, by a germanium resistor built into the pressure cell near the sample position.

3.  **Results and Discussion**

The magnetic and transport properties of the $RuSr_2GdCu_2O_8$ sample chosen for high pressure measurements are similar to that previously reported by different groups. Figure 1 shows the dc susceptibility, $\chi_{dc}$, as a function of temperature. The increase of $\chi_{dc}$ at about 130 K clearly indicates the onset of the ferromagnetic order. A strong diamagnetic signal appears in the zero field cooled (zfc) susceptibility below 30 K. The inset of Fig. 1, however, shows that superconductivity actually sets in at a higher temperature of $T_c = 42$ K. The small diamagnetic signal below 42 K is assigned to the intra-grain superconductivity whereas the larger signal below 30 K is due to the inter-grain shielding signal.

The real part of the ac-susceptibility, $\chi'_{ac}$, exhibits features very similar to the zfc dc-susceptibility, a well defined peak at $T_m$ and the two steps in the diamagnetic drop indicating the intra- and inter-grain superconducting transitions (Fig. 2). Under high pressure $\chi'_{ac}$ is easily measured using the mutual inductance method.

The ambient pressure resistivity, $\rho$, and thermoelectric power, S, are shown in Fig. 3A. At the ferromagnetic transition $\rho$ shows a small but distinct change of slope as shown in more detail in the upper left inset of the figure. The Seebeck coefficient is positive and large at room temperature (70 µV/K) and its temperature dependence is typical for an under-doped high-$T_c$ superconductor. The superconducting transition proceeds in two steps, according to the intra-grain and inter-grain transitions. This is clearly seen in the derivative, $d\rho/dT$, showing two well resolved maxima (Figure 3B). The position of these maxima can be resolved by fitting two Gaussian shaped peaks to the data of Fig. 3B. The center positions of the peaks are then used to define the intra-grain ($T_c$) and inter-grain ($T_p$) transition temperatures from the resistivity measurements.

At high pressures both quantities, $\chi'_{ac}$ and $\rho$, are measured simultaneously and the estimated critical temperatures as a function of pressure are shown in Figure 4. Both temperatures, $T_c$ and $T_p$, increase linearly with p but at different rates. The open symbols denote the $T_c$'s as derived from the onset of the diamagnetic drop of $\chi'_{ac}$ (triangles) and from the high temperature peak of $d\rho/dT$ (circles). The small difference of the absolute value of both $T_c$'s is due to their definitions. The pressure shifts of $T_c$ of 1.02 K/GPa and 1.06 K/GPa obtained from $\chi'_{ac}$ and $\rho$, respectively, are consistent. The larger pressure shift of the inter-grain phase lock temperature $dT_p/dp=1.8$ K/GPa is a consequence of the pressure-induced improvement of the grain-grain contacts resulting in an additional enhancement of the inter-grain Josephson coupling.

The ferromagnetic transition temperature is also found to increase with pressure at a linear rate of 6.7 K/GPa (Fig. 5). This rate is distinctly larger than that of the superconducting $T_c$. Comparing the relative pressure coefficients, $d\ln T_c/dp=0.025$ and $d\ln T_m/dp=0.054$, the magnetic $T_m$ increases still about twice as fast with p as the superconducting $T_c$. The ferromagnetic state appears to be strongly stabilized under pressure which should have some consequences for the superconducting state.

In fact, the pressure coefficient of the superconducting $T_c$ of $RuSr_2GdCu_2O_8$ is by a factor of 3 to 4 smaller than that of other under-doped high-$T_c$ compounds, e.g. $La_{2-x}(Sr,Ba)_xCuO_4$ [21], $YBa_2Cu_3O_{7-\delta}$ [20], or $YBa_2Cu_{3-x}M_xO_{7-\delta}$ [22]. In particular the comparison with the iso-structural $YBa_2Cu_3O_{7-\delta}$ suggests that the magnetic order in $RuSr_2GdCu_2O_8$ possibly reduces the expected enhancement of $T_c$ under pressure. The small pressure coefficient of 1 K/GPa is then an immediate consequence of a competition of ferromagnetic and superconducting phases. Due to this competition the stronger enhancement of the magnetic phase results in a reduced (as compared to YBCO) pressure effect on $T_c$. It would be interesting to compare the pressure effects of the stoichiometric $RuSr_2GdCu_2O_8$ with that of the Cu doped $Ru_{1-x}Sr_2GdCu_{2+x}O_8$ where it was recently shown that the magnetic order is largely suppressed [19].

4. Conclusions

We have investigated the effect of hydrostatic pressure on the intra-grain superconducting and ferromagnetic transitions in $RuSr_2GdCu_2O_8$. Although many physical properties of this compound are very similar to the under-doped $YBa_2Cu_3O_{7-\delta}$ the pressure coefficient of the superconducting $T_c$ was found to be unusually small. The relative positive pressure effect on the magnetic phase is about twice as large as that on the sc phase. The current data suggest the existence of a strong competition between superconducting and magnetic states in the superconducting ferromagnets.

**Acknowledgements**

This work is supported in part by NSF Grant No. DMR-9804325, the T.L.L. Temple Foundation, the John J. and Rebecca Moores Endowment, and the State of Texas through the Texas Center for Superconductivity at the University of Houston and at Lawrence Berkeley Laboratory by the Director, Office of Energy Research, Office of Basic Energy Sciences, Division of Materials Sciences of the U.S. Department of Energy under Contract No. DE-AC03-76SF00098.

Figure Captions

Fig. 1: dc-susceptibility of $RuSr_2GdCu_2O_8$. Upper and lower branches denote field cooled and zero field cooled data, respectively, measured at 7 Oe. The inset shows the details at the superconducting transition.

Fig. 2: Real part of the ac-susceptibility of $RuSr_2GdCu_2O_8$. $T_m$ denotes the ferromagnetic transition temperature. The inset shows an enlarged section close to the superconducting transition and the definitions of $T_c$ and $T_p$.

Fig. 3A: Resistivity and thermoelectric power of $RuSr_2GdCu_2O_8$. The uper left inset shows the change of resistivity slope at the ferromagnetic transition.

Fig. 3B: Derivative $d\rho/dT$ at the superconducting transition. The two peaks assigned to intra- and inter-grain transitions are well resolved and shown by dashed lines.

Fig. 4: Pressure dependence of $T_c$ estimated from $\chi'_{ac}$ (open triangles) and $\rho$ (open circles) and $T_p$ (filled circles).

Fig. 5: Pressure dependence of the ferromagnetic transition temperature, $T_m$.

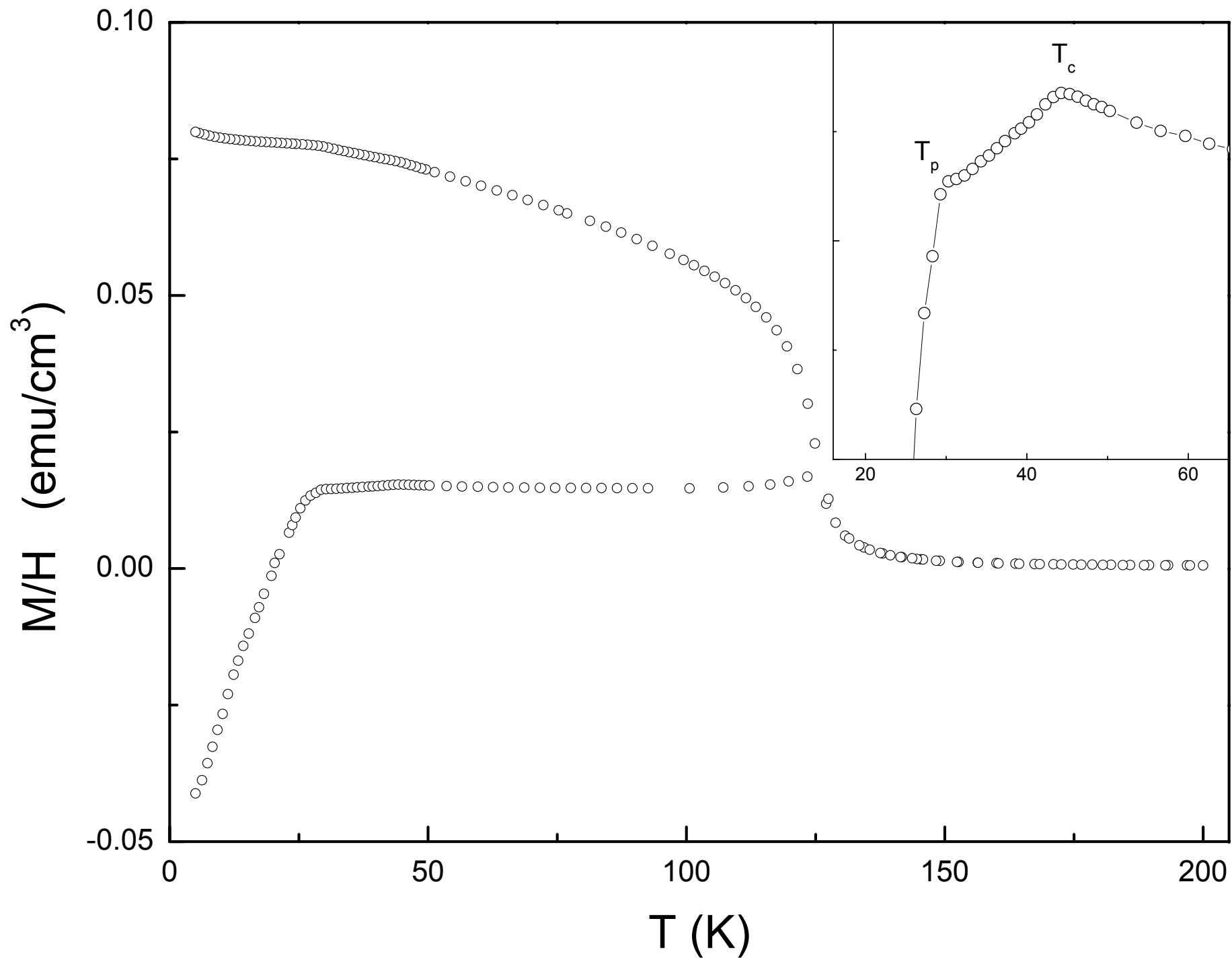

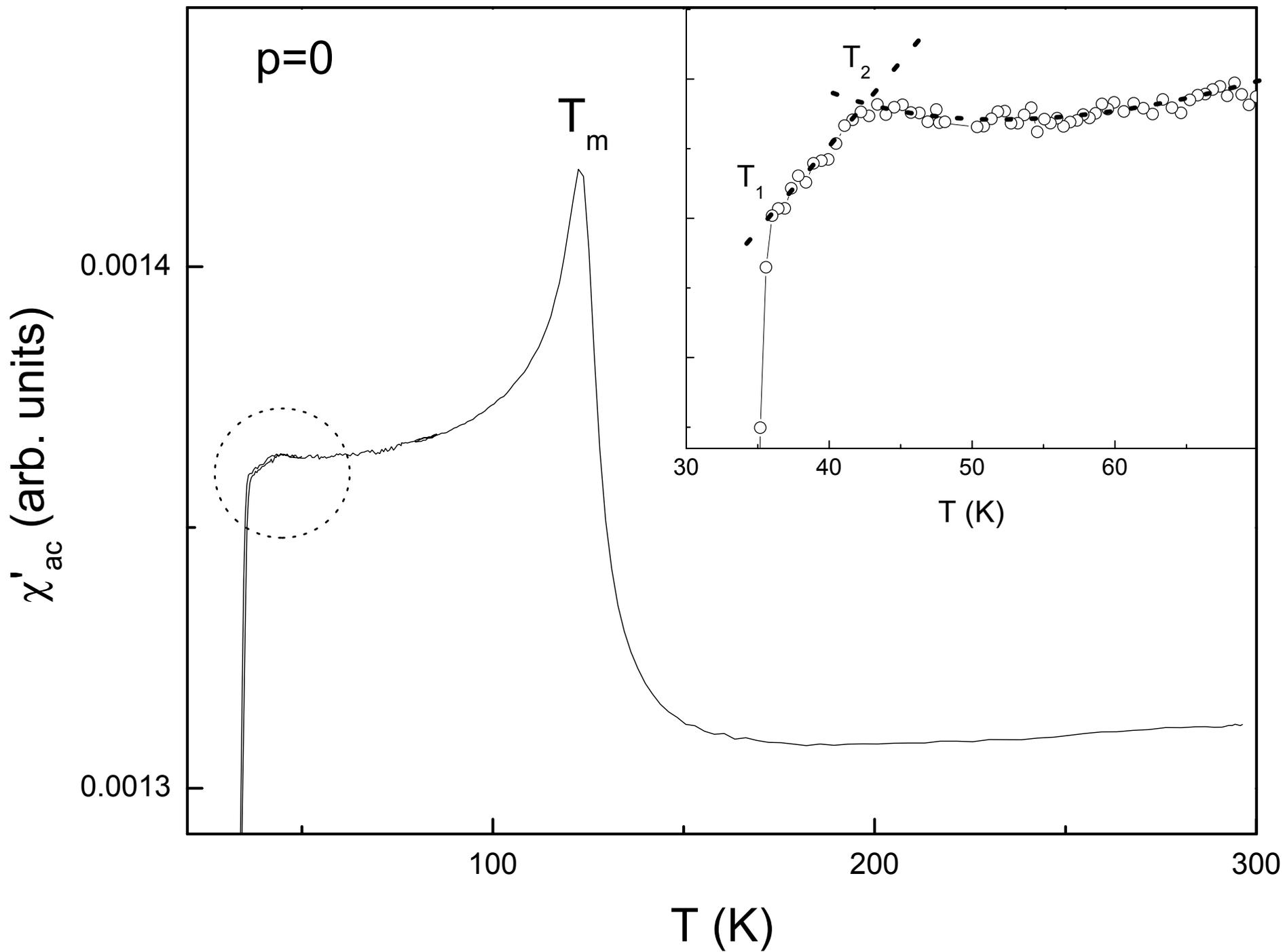

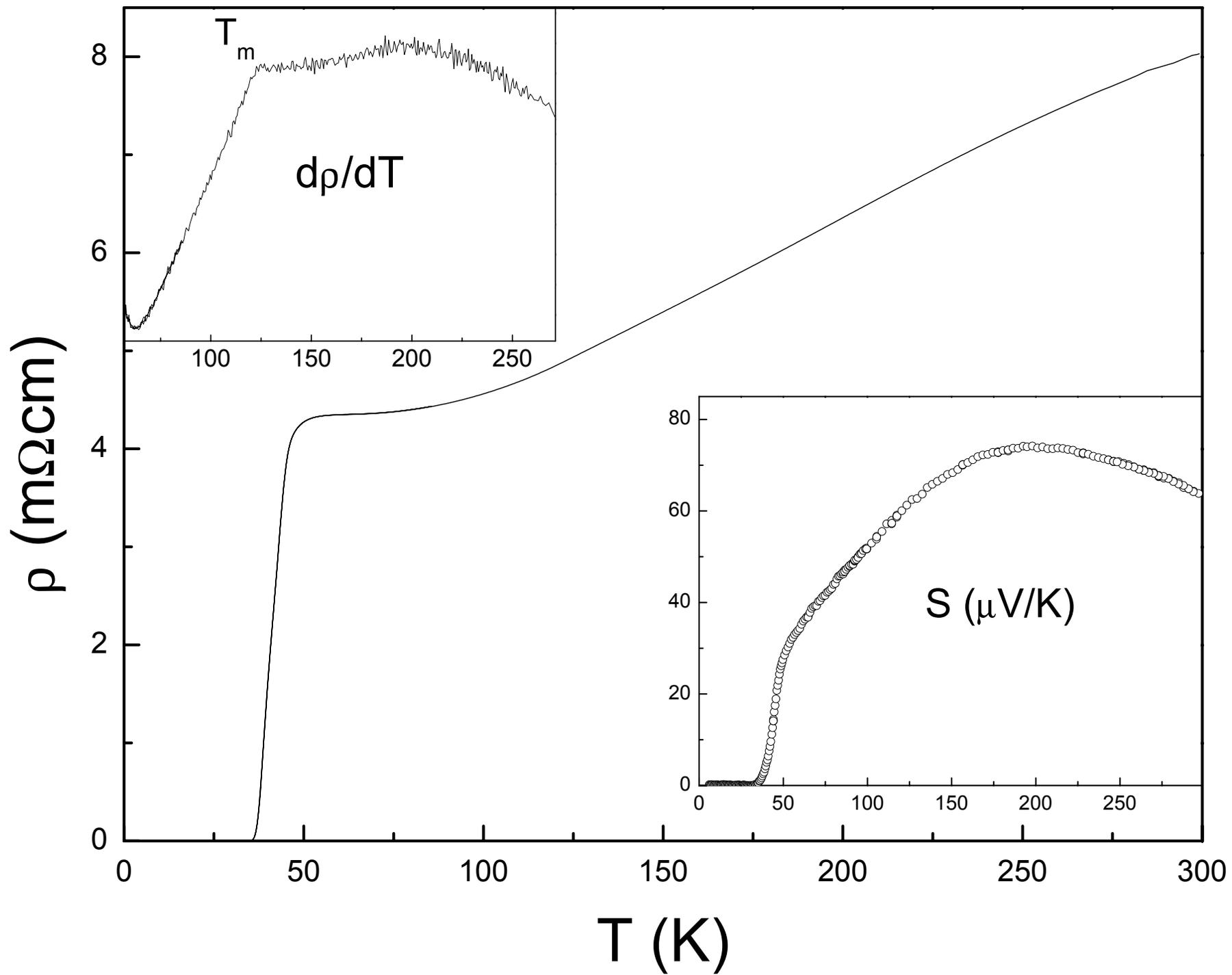

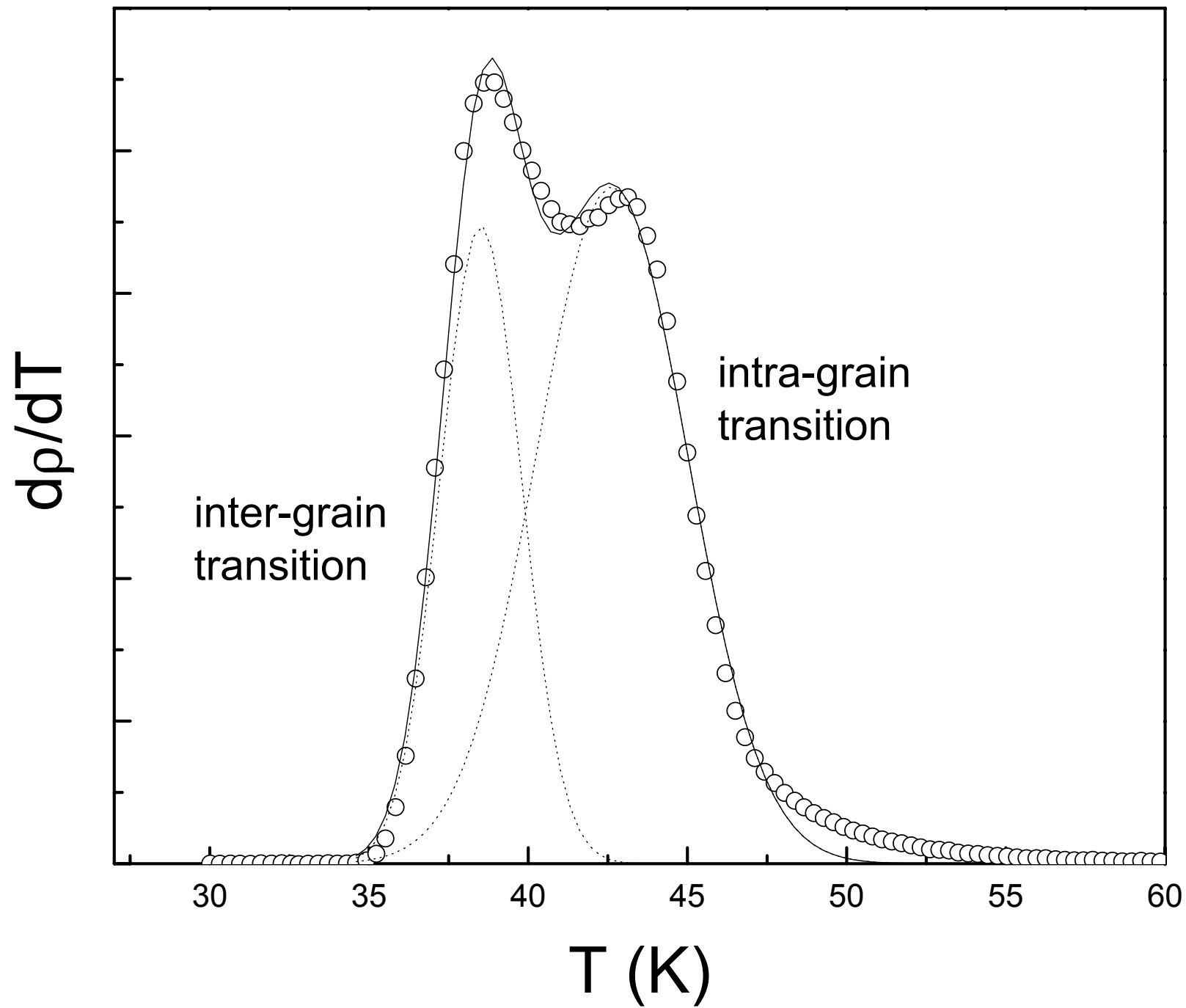

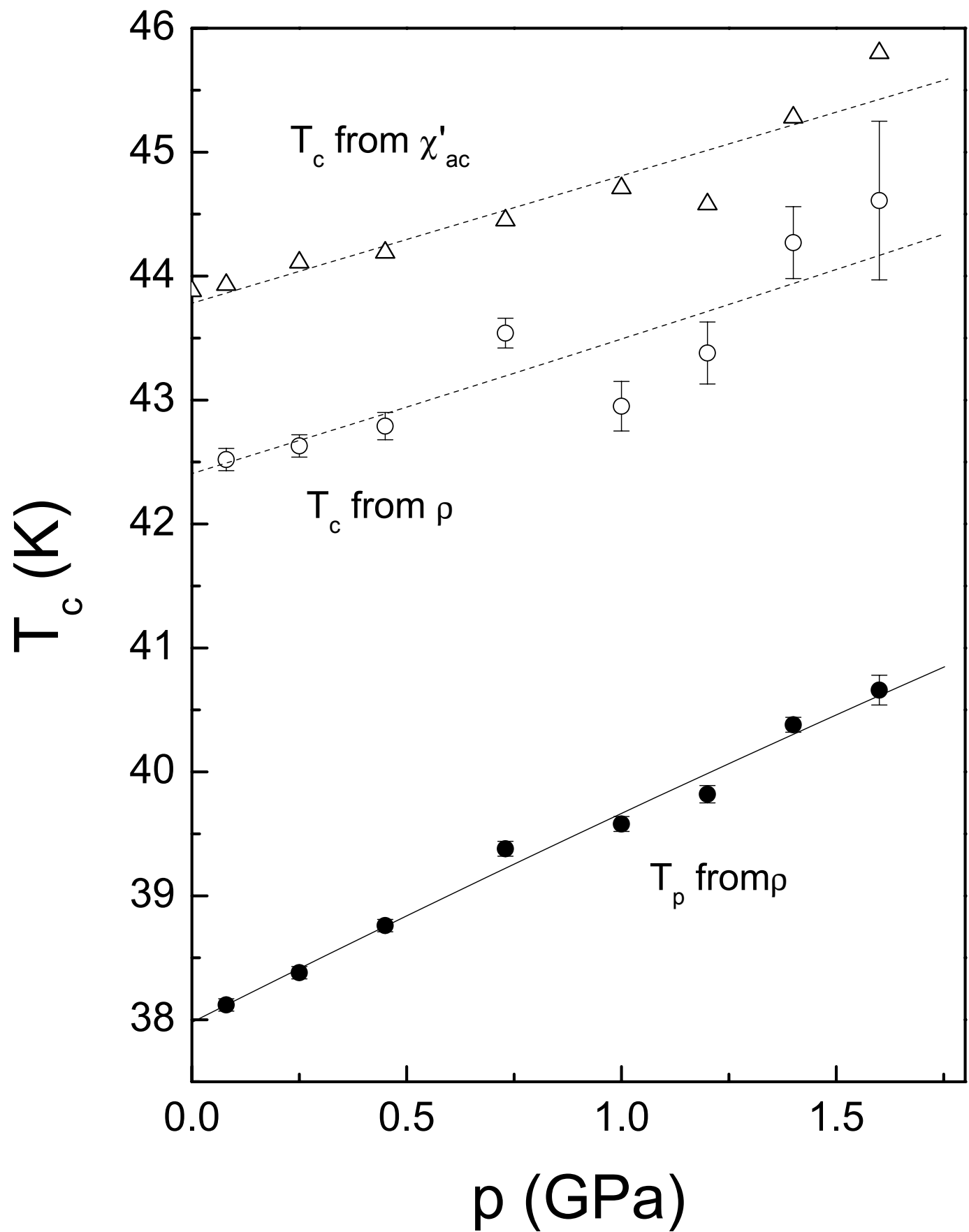

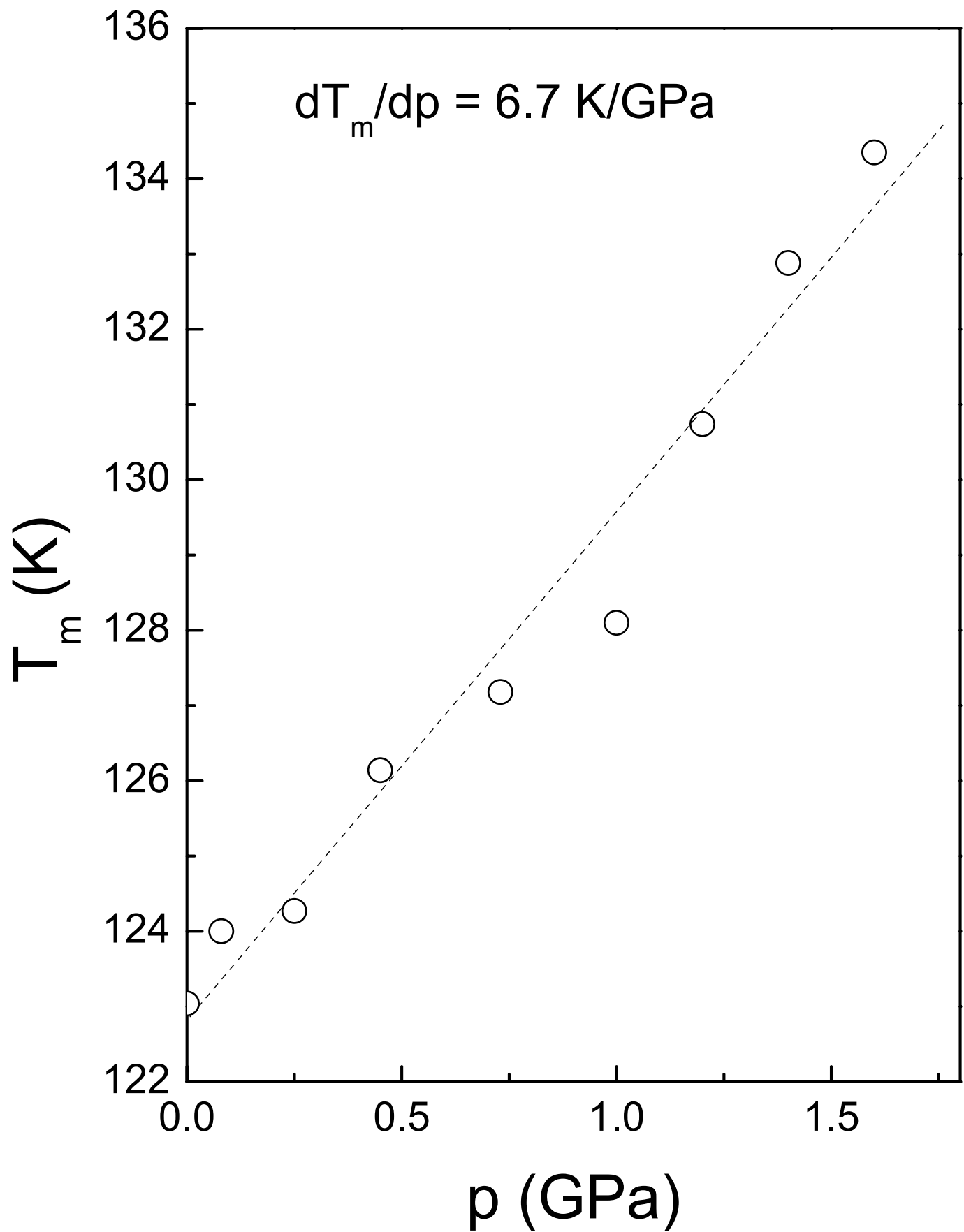